\documentclass[manuscript]{./aastex}

\def\lax{{$\mathrel{\hbox{\rlap{\hbox{\lower4pt\hbox{$ \sim $}}}\hbox{$<$}}}$}}
\def\gax{{$\mathrel{\hbox{\rlap{\hbox{\lower4pt\hbox{$ \sim $}}}\hbox{$>$}}}$}} 

\usepackage{color}

\newcommand{\fmin}{f_\mathrm{min}}
\newcommand{\fmax}{f_\mathrm{max}}
\newcommand{\flow}{f_\mathrm{low}}
\newcommand{\fhigh}{f_\mathrm{high}}

\newcommand{\fisco}{f_\mathrm{ISCO}}

\slugcomment{}

\shorttitle{Gravitational wave diagnosis of a circumbinary disk}
\shortauthors{Hayasaki, Yagi, Tanaka \& Mineshige}

\begin{document}

\title{Gravitational wave diagnosis of a circumbinary disk}
\author{Kimitake Hayasaki}
\email{kimi@kusastro.kyoto-u.ac.jp}
\affil{Department of Astronomy, Kyoto University, Oiwake-cho, Kitashirakawa, Sakyo-ku, Kyoto 606-8502, Japan}
\author{Kent Yagi}
\affil{Department of Physics, Kyoto University, Oiwake-cho, Kitashirakawa, Sakyo-ku, Kyoto 606-8502, Japan}
\author{Takahiro Tanaka}
\affil{Yukawa Institute for Theoretical Physics, Kyoto University, Oiwake-cho, Kitashirakawa, Sakyo-ku, Kyoto 606-8502, Japan}
\author{Shin Mineshige}
\affil{Department of Astronomy, Kyoto University, Oiwake-cho, Kitashirakawa, Sakyo-ku, Kyoto 606-8502, Japan}

%
%

\begin{abstract}
When binary black holes are embedded in a gaseous environment, 
a rotating disk surrounding them, the so-called circumbinary disk, 
will be formed. The binary exerts a gravitational torque on the 
circumbinary disk and thereby the orbital angular momentum is 
transferred to it, while the angular momentum of the circumbinary 
disk is transferred to the binary through the mass accretion.
The binary undergoes an orbital decay due to both the gravitational 
wave emission and the binary-disk interaction. This causes the 
phase evolution of the gravitational wave signal. The precise 
measurement of the gravitational wave phase thus may provide 
information regarding the circumbinary disk. In this paper, we 
assess the detectability of the signature of the binary-disk interaction
 using the future space-borne gravitational wave detectors such as 
 {\it DECIGO} and {\it BBO} by the standard matched filtering analysis. 
 We find that the effect of the circumbinary disk around binary black 
 holes in the mass range $6M_\odot\le{M}\lesssim3\times10^3M_\odot$ 
 is detectable at a statistically significant level in five year observation, 
 provided that gas accretes onto the binary at a rate greater than 
 $\dot{\it M}\sim1.4\times10^{17}\,[\rm{g\,s^{-1}}]\,{\it j}^{-1}({\it M}/10{\it M}_\odot)^{33/23}$ 
 with $10\%$ mass-to-energy conversion efficiency, where $j$ represents 
 the efficiency of the angular momentum transfer from the binary to 
 the circumbinary disk. We show that $O(0.1)$ 
 coalescence events are expected to occur in sufficiently dense 
 molecular clouds in five year observation. We also point out that the 
 circumbinary disk is detectable, even if its mass at around the inner 
 edge is by over $10$ orders of magnitude less than the binary mass.
\end{abstract}

\keywords{black hole physics - accretion, accretion disks - gravitational waves - stars: evolution}
%
\section{Introduction}
\label{sec:intro}
%

It is widely accepted that coalescing binary compact objects, 
{\it i.e}., those consisting of black holes, neutron stars, or white dwarfs,
are important sources of gravitational waves (GWs). 
Ground-based detectors with higher sensitivity 
are currently under construction, {\it e.g}.,
Advanced-LIGO, Advanced-VIRGO, and LCGT,
while there are plans for the space-borne interferometers 
such as {\it LISA} \citep{d97}, {\it DECIGO} \citep{seto01} 
and {\it BBO} \citep{p03}. The significant improvement 
of sensitivity will make it possible to directly 
detect GWs in near future. 
{\it LISA} has its best sensitivity at around $10^{-3}-10^{-2}$\,Hz and one of its main targets is a merger event of binary massive black holes. On the other hand, {\it DECIGO/BBO} has its best sensitivity at around $0.1-1$\,Hz. 
Its ultimate goal is to detect primordial GW backgrounds, but it has promising astrophysical sources such as binary stellar-mass black holes. It is also sensitive to merging signals from binary intermediate-mass black holes.

Recently, much attention has been paid to the surrounding 
environment of the GW sources. 
Several types of possible electromagnetic (EM) signatures associated with GW emissions 
have been proposed, such as afterglows \citep{mp05}, precursors 
\citep{cp09,bode10} and periodic emission \citep{hms07,bo08,mm08,cu09} 
in the context of a massive black hole coalescence.
Those EM signatures will be also very useful in localizing 
the position of a GW source in the sky and in identifying its redshift.
By combining the redshift information 
with a luminosity distance determined by the GW signal, 
one can obtain a distance-redshift relation.
This will be one of the best observational probes of the 
dark energy \citep{schutz86, hh05}. 
However, there are not so many works discussing EM signals  
from coalescing binary black holes in the stellar/intermediate 
mass range $6-10^4M_\odot$.
Obviously, no black holes can emit EM radiation, 
unless they interact with their environmental gas. 
Thus, our interest is naturally led to 
binary stellar/interstellar-mass black holes 
in a dense gaseous medium. 

\cite{nr00} investigated the effect of hydrodynamic interactions 
between an advection-dominated accretion flow 
onto a supermassive black hole and inspiraling 
stars on GW signals.
This effect was shown to be negligible 
as long as the advection-dominated accretion flow is concerned.
The effects of the interaction with a gaseous disk 
on the GW phase evolution have been also discussed 
in the context of an extreme mass ratio inspiral (EMRI), 
where a compact object inspirals 
into a supermassive black hole \citep{br08,kocsis11,yunes11}.
In this case the situation is more similar to a planet migration.
It has been claimed that the existence of an accretion disk 
around a supermassive black hole is detectable for a certain parameter range with {\it LISA}.

In contrast, a binary composed 
of black holes with nearly equal masses residing in a gaseous environment 
will form a triple disk system: two accretion disks around the respective  
black holes and one circumbinary disk surrounding both of them \citep{hms07,hmh08}.
Then, the binary black holes and the circumbinary disk should mutually interact,
which inevitably affects the orbital evolution of the binary. 
The gas surrounding the binary acts on the orbital motion 
through a drag force (tidal/resonant interaction), 
leading to the angular momentum transfer from the binary to the gas.
Conversely, the angular momentum is carried from the circumbinary disk to the binary by gas accretion. 
Since {\it DECIGO/BBO} will detect many cycles of GWs from  
the binary inspiral, a rather tiny correction in the GW phase 
is detectable.
Therefore signature of the circumbinary disk can  
arise in the GW signal even if the circumbinary disk is extremely less massive by many
orders of magnitude than the binary.
In this paper, we give an estimate how large the mass accretion rates 
of the circumbinary disk should be, so that we can 
detect the effect of the binary-disk interaction on 
the GW phase evolution at a statistically significant 
level with {\it DECIGO/BBO}.
Note that we evaluate the effect of the binary-disk interaction 
at the first order in a mass accretion rate of the circumbinary disk 
throughout this paper.
Although our current understanding of the binary-disk interaction is poor, 
we assume here that we knew how the GW phase evolution is modified and 
hence the standard matched filtering analysis is applicable. 
Based on our estimate of the mass accretion rate required for detection,
we also give a prediction for the number of merger events under 
detectable influence of the circumbinary disk.
%
%
\begin{figure}[!ht]
\resizebox{\hsize}{!}{
\includegraphics[angle=-90, width=80mm]{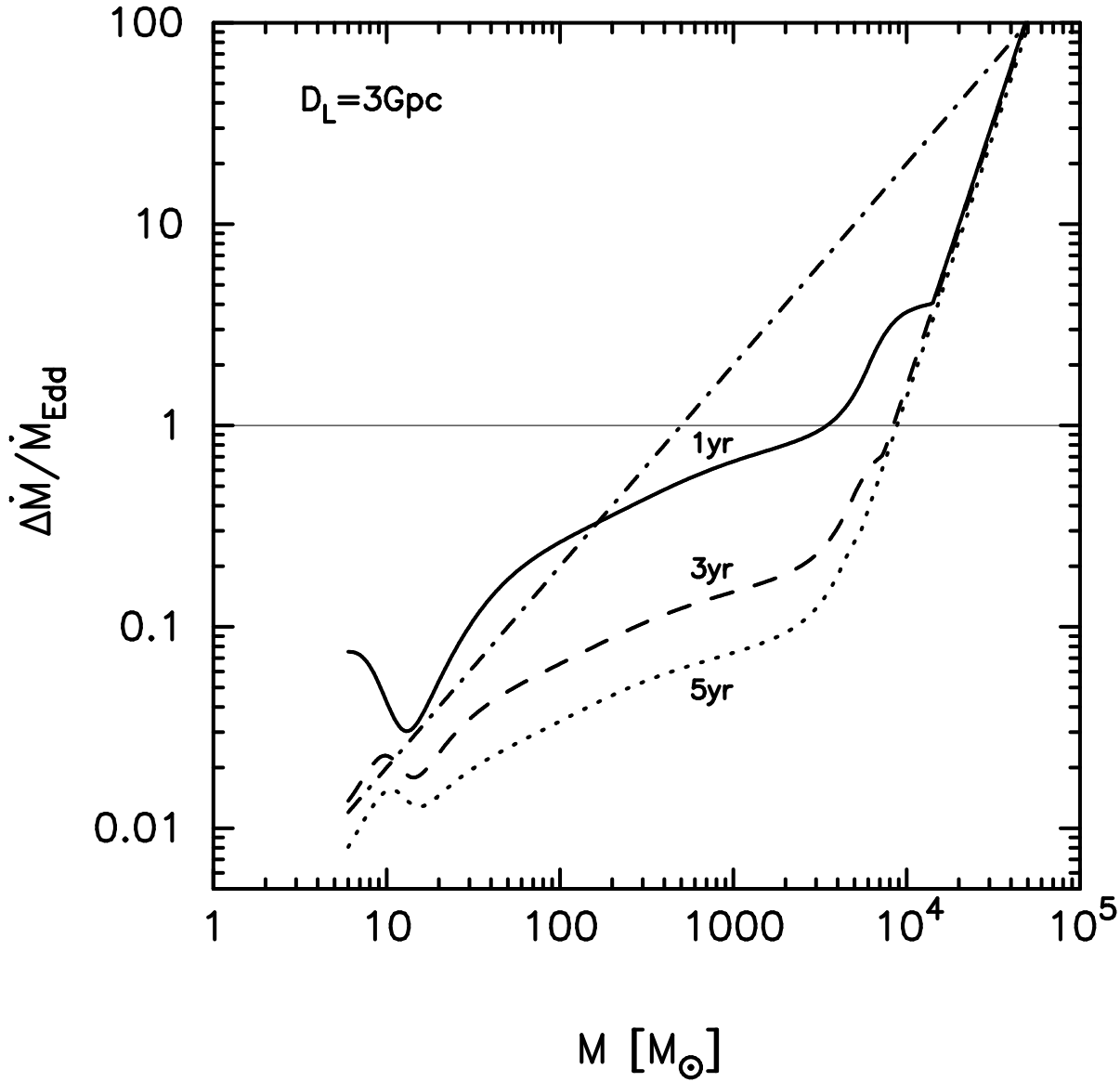}
}
\caption{
Determination accuracy $\Delta\dot{M}$ of mass accretion 
rates of the circumbinary disk around coalescing binary black holes 
within $D_{L}=3\,\rm{Gpc}$. Here, $j=1$ is adopted 
(see equation~(\ref{dotJoverJ}) for the definition of $j$).
The determination accuracy 
is normalized by the Eddington accretion rate with 
$10\%$ mass-to-energy conversion efficiency.
$M$ is the total mass of binary black holes 
in units of the solar mass. 
The solid, dashed and dotted curves correspond to 
the determination accuracies with the observation time, 
$t_0-t_{\rm{ini}}=$1, 3, and 5\,yr, respectively.
The dash-dotted line shows the Bondi-Hoyle-Lyttelton accretion rate
normalized by the Eddington rate, where the BHL rate is given by 
equation~(\ref{bhlrate}).
The mass accretion rate can be measured from GW observations 
at a statistically significant level if $\Delta{\dot{M}}$ 
is less than the fiducial accretion rate $\dot{M}$.
The horizontal solid thin line represents $\Delta{\dot{M}}=\dot{M}_{\rm{Edd}}$.
}
\label{fig:da}
\end{figure}

\begin{figure}[!ht]
\resizebox{\hsize}{!}{
\includegraphics[angle=-90, width=80mm]{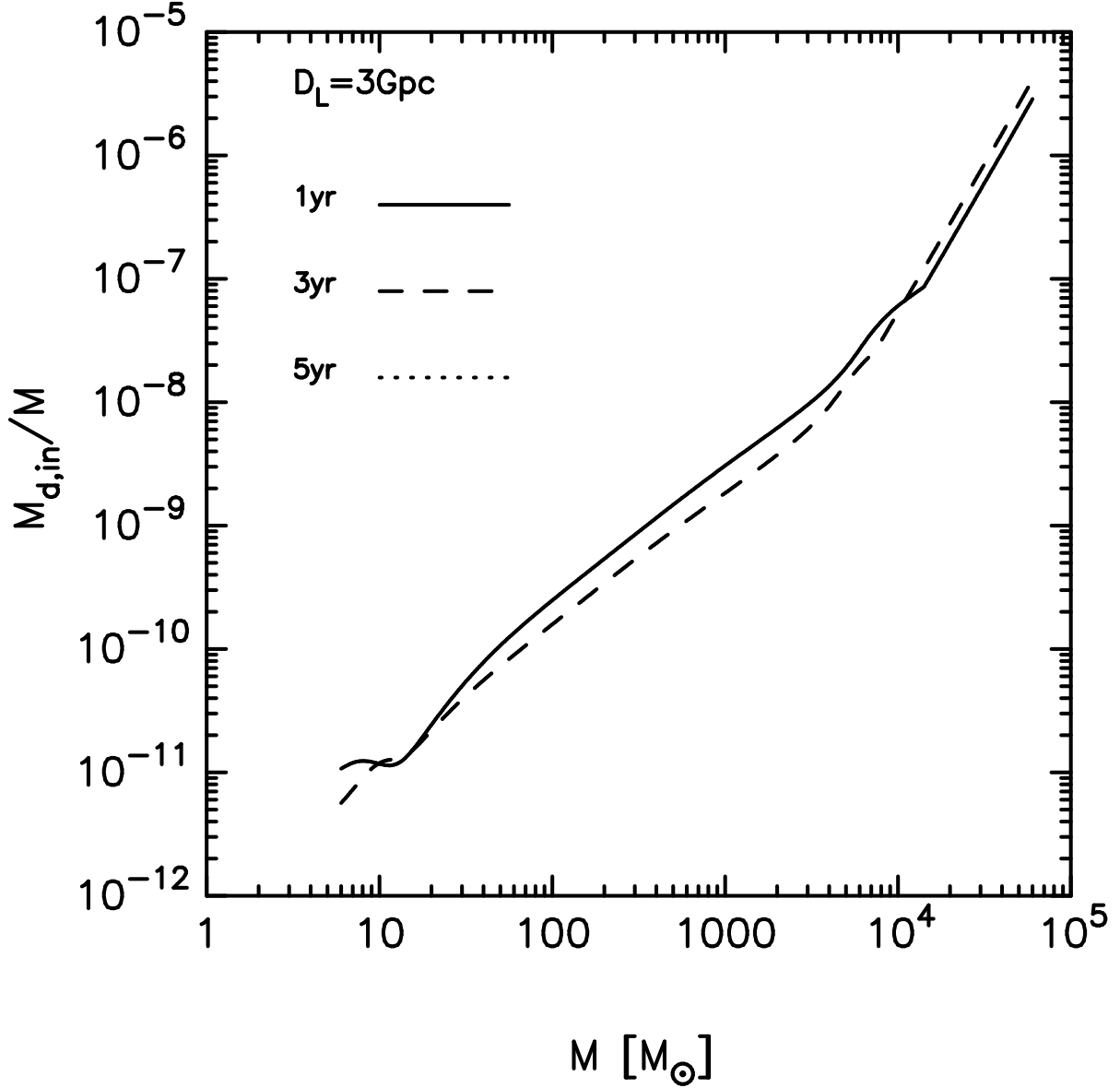}
}
\caption{
The inner-most disk mass that gives detectable effect on the 
GW phase evolution is plotted as a function of the total mass of binary black holes. 
The vertical axis is normalized by the total mass of binary black holes, and 
the inner-most disk mass is defined by 
$M_{\rm{d,in}}=\pi{r_{\rm{in}}^2}\Sigma(r_{\rm{in}})$,
where $r_{\rm{in}}$ and $\Sigma$ are 
the inner edge radius of circumbinary disk and the surface density 
of the circumbinary disk evaluated at around $r_{\rm{in}}$, respectively
(see equation~(\ref{eq:dm}) in the text).
The solid, dashed and dotted curves correspond to 
$M_{\rm{d,in}}/M$ for $t_0-t_{\rm{ini}}=$1, 3, and 5\,yr, respectively.
}
\label{fig:dm}
\end{figure}
%
\section{Measuring mass accretion rates with GW detectors}
\label{sec:da}
%
In this section, we first briefly explain the basics 
about how the phase of GWs from a binary evolves 
as a result of the radiation reaction due to the GW 
emission. Second, we present the picture that 
we have in mind about the interaction 
between the binary and the circumbinary disk, 
providing a quantitative estimate for the correction 
due to the binary-disk interaction in the GW phase evolution.
We relate the magnitude of the correction to the mass accretion rate 
of the circumbinary disk. Then, we evaluate the 
determination accuracy of the mass accretion rate of the circumbinary disk.
%
\subsection{Standard gravitational wave phase evolution}
\label{subsec:ini}
%
Let us first briefly review the evolution of the GW phase of 
a coalescing binary, neglecting the interaction between the binary and the circumbinary disk.
Close enough binaries are expected to evolve their orbits by losing 
the energy and angular momentum due to the GW emission. 
Here, we assume that binary black holes are in a circular orbit, 
because the orbital eccentricity decreases rapidly in proportion to $a^{19/12}$ \citep{p64},
where $a$ is the semi-major axis of the binary. 
At the leading order of the Post-Newtonian (PN) expansion, 
the binding energy and orbital angular momentum of a 
circular binary are, respectively, given by
\begin{eqnarray}
\mathcal{E}
=
-\frac{\mu c^2}{2}x,
\qquad
\mathcal{J}
=\frac{\mu c^2}{\Omega_{\rm orb}}x\,,
\label{binde}
\end{eqnarray} 
where $c$ is the speed of light, 
$\mu\equiv M_1 M_2/M$ is the reduced mass with $M\equiv M_1+M_2$ representing the total mass
of binary black holes, $x=({GM}\Omega_{\rm orb}/c^3)^{2/3}$ is a non-dimensional quantity of
$O((v/c)^2)$, and $\Omega_{\rm orb}=\sqrt{GM/a^3}$ is an orbital frequency of the binary. 
The leading terms in the GW energy and angular momentum fluxes emitted
from a binary are given by the well-known quadrupole formulas as
\begin{eqnarray}
\dot \mathcal{E}_{\rm{gw}}
=
\frac{32c^5}{5G}
\eta^2
x^5\, ,
\qquad
\dot \mathcal{J}_{\rm{gw}}
=
\frac{32c^5}{5G\Omega_{\rm orb}}
\eta^2
x^5\,
\label{gwflux}
\end{eqnarray}
with $\eta\equiv\mu/M$ and a dot represents differentiation with respect to time. 
For an equal-mass binary, $\eta$ is set to be $0.25$.
The orbit of the binary gradually decays as a result of the loss 
of its energy and angular momentum due to the GW emission. 
Equating the time variation of the orbital energy or angular momentum 
with the loss due to the GW emission, 
one can derive, to the lowest PN order,  
$\dot{x}= 64c^3\eta x^5/5GM$, which is translated into the evolution 
equation for the GW frequency $f=\Omega_{\rm orb}/\pi$ as  
\begin{eqnarray}
\dot{f}
=
\frac{96\pi}{5}
f^2
\left(\frac{\pi{G}\mathcal{M}f }{c^3}\right)^{5/3}, 
\label{quadrupoledotf}
\end{eqnarray}
where we introduced the chirp mass, $\mathcal{M}\equiv \eta^{3/5}M$. 

Integrating this equation with the coalescence time set to $t_0$, 
the time evolution of the GW frequency is computed as 
\begin{eqnarray}
f(t)
&=& {c^3\over \pi GM}\left(\frac{256}{5}\frac{c^3(t_0-t)}{GM}\eta\right)^{-3/8}
\cr
&\sim& 
9.3\times10^{-2}[\rm{Hz}]\,
\eta_{0.25}^{-3/8}
{\it m}_{10}^{-5/8}
\left(\frac{\it t_{\rm{0}}-t}{1yr}\right)^{-3/8}
\,,
\label{fobs}
\end{eqnarray}
where we use $\eta_{0.25}\equiv\eta/0.25$ and $m_{10}\equiv M/10M_\odot$ for brevity.
One can translate this frequency evolution into the evolution of $a$,
using the relation  $(\pi f)^2=GM/a^3$, that is,
\begin{eqnarray}
a(t)
&=&\frac{GM}{c^2}\left(\frac{256}{5}\frac{c^3(t_0-t)}{GM}\eta\right)^{1/4}
\\
&\sim&2.5\times10^9[\rm{cm}]\,
\eta_{0.25}^{1/4}
{\it m}_{10}^{3/4}
\left(\frac{\it t_{\rm{0}} - t}{1yr}\right)^{1/4}\,.
\label{acoales}
\end{eqnarray} 
As the orbit shrinks  with time, the GW frequency gradually increases
and finally reaches the frequency at the inner-most stable circular orbit (ISCO)
\begin{eqnarray}
f_{\rm{ISCO}}
=\frac{c^3}{6^{3/2}\pi GM}\sim440[\rm{Hz}]\,
{\it m}_{10}^{-1}\,,
\label{fisco}
\end{eqnarray}
which is here approximated by the expression 
in the EMRI limit irrespective of the mass ratio. 

The number of the orbital cycles before the coalescence 
is then calculated as
\begin{eqnarray}
N_{\rm{cyc}}
&\equiv&
\int^{t_0}_{t}\frac{f(t)}{2}dt
\nonumber \\
&\sim&
2.3\times10^6
\,
\eta_{0.25}^{-3/8}
m_{10}^{-5/8}
\left(\frac{\it t_{\rm{0}}-t}{1\rm{yr}}\right)^{5/8}.
\end{eqnarray}
Our main interest is in binary black holes within the mass range  
$10-10^4M_\odot$. When the mass of each black hole is 
$10M_\odot$ and the observation starts one year before the coalescence, 
the initial GW frequency and the initial semi-major axis are 
$\sim0.1\rm{Hz}$ and $\sim0.04R_\odot$, respectively. 
In the typical case one can observe 
$N_{\rm{cyc}}\sim 10^6$ cycles of GWs. Such a 
large number of orbital cycles allow us to  
precisely measure the GW phase correction possibly 
to the level sensitive to the tiny effect of the circumbinary disk.

The observed waveform of GWs from the binary 
at the leading PN order 
is given by \citep{cf94}
\begin{eqnarray}
h(t)=
\left(\frac{384}{5}\right)^{1/2}
\pi^{2/3}
Q
\frac{(GM)^2}{a(t)D_{L}c^4}
\eta
\cos\phi(t)\,,
\label{ht}
\end{eqnarray}
where 
$Q$ is a factor of at most order unity 
that depends on the direction and orientation of 
the binary 
and $D_{L}$ is the distance from the earth to the GW source. 
The GW phase $\phi(t)$ is defined by 
\begin{eqnarray}
\phi(t)
=
\phi_0
+
2\pi
\int_{t_0}^{t} fdt\,,
\label{gwp}
\end{eqnarray}
where $\phi_0$ is the GW phase at the coalescence. 
%
\subsection{Correction to the GW phase due to binary-disk interaction}
\label{subsec:da}
%
In this subsection, we consider the correction to the phase evolution of
GWs emitted 
from an inspiralling binary caused by the interaction with the circumbinary disk. 
Such binary-disk interaction induces additional extraction of the orbital angular
momentum from the binary. 
This effect will be tiny compared to the phase evolution driven by 
the GW emission when the binary separation is close 
enough to be observable by future GW detectors. 
Even in that case, however, the correction might 
be observable because of a large number of cycles of GWs 
in the frequency range of observations.
For simplicity, we assume that the orbital evolution
is dominated by the radiation reaction due to the GW emission. 
In this case it would be natural to assume that the 
orbit is well approximated by a quasi-circular one. 
Although the circumbinary disk is not necessarily aligned to the 
binary orbital plane, here we also assume a simple aligned configuration. 

The exchange between the orbital angular momentum of the binary,
$\mathcal{J}$, and the angular momentum of the circumbinary disk 
is in two ways. One is the angular momentum transfer 
through the tidal/resonant interaction \citep{al94,armi1}.
The other is through the overflow of the gas from the 
inner edge of circumbinary disk onto the central binary \citep{al96,hms07}. 
The transferred gas will form accretion disks around 
respective black holes~\citep{hmh08}. 
During this accretion process, some fraction of the angular momentum of the transferred gas will
temporally go to the accretion disks associated with respective black holes.
However, since the specific angular momentum of gas at the inner edges of
the accretion disks is so tiny, only little fraction of the angular momentum of
the accretion disks is transferred to the black hole spins. 
Thus, most of this angular momentum is 
transferred outward via the viscous process and 
is finally added to the orbital angular momentum of the binary 
$\mathcal{J}$ via the tidal interaction~\citep{kh09}.
Therefore we can neglect the angular momentum that ends up with 
the spins of the black holes. Thus, the net angular momentum extracted
from the binary orbital one due to interaction with the circumbinary
disk, $\dot{\cal J}_{\rm disk}$, should be almost identical to the 
net angular momentum flow inside the circumbinary disk, 
$\dot{J}_{\rm{disk}}$, which simply follows from the conservation law. 
(We take the flow of the angular momentum to be positive when it flows
outward.)

Now, we roughly estimate the order of magnitude of
$\dot{J}_{\rm{disk}}$, 
without relying on a specific model of the circumbinary disk.
For a quasi-stationary disk, 
$\dot{J}_{\rm{disk}}$ should be independent of the radius $r$, 
which should 
be equal to the sum of 
the flows of angular momentum by the viscous torque $\dot{J}_{\rm{vis}}\,(>0)$ 
and by the mass accretion $\dot{J}_{\rm{acc}}\,(<0)$, 
\begin{equation} 
\dot{J}_{\rm{disk}}
=\dot{J}_{\rm{vis}}(r)+\dot{J}_{\rm{acc}}(r)
=\rm{const.}\,, 
\label{Jdotdr}
\end{equation}
except near the inner edge of the disk located at $r=r_{\rm{in}}$, 
where the tidal/resonant torque is non-negligible. 
Note that $\dot{J}_{\rm{acc}}(r)$ is negative when the mass 
is accreting inward.
Since $|\dot{J}_{\rm{acc}}(r)|$ increases with $r$ like $\propto\sqrt{r}$ for a 
Keplerian disk, 
$\dot{J}_{\rm{vis}}(r)$ and $|\dot{J}_{\rm{acc}}(r)|$
approximately balance with each other except near the 
inner edge of the circumbinary disk. 
If we were discussing an accretion disk around a single central object, the radius 
of the inner edge of the disk 
would be much smaller than that of the circumbinary disk.
In this case, $\dot{J}_{\rm{vis}}(r)$ should be almost balanced with 
$|\dot{J}_{\rm{acc}}(r)|$ near $r=r_{\rm{in}}$.
In the present situation, however, 
the binary exerts the tidal/resonant torque, which 
truncates the disk at $r=r_{\rm{in}}$. 
Then, the surface density of the disk is enhanced near the
inner edge by a factor of two or so compared with the case without the
tidal/resonant torque~(e.g., see Figure~3 of \citealt{hms07}).
Accordingly, the viscous torque is also locally enhanced by about the
same ratio. 
This implies that $\dot{J}_{\rm{disk}}\approx
|\dot{J}_{\rm{acc}}(r_{\rm{in}})|$.
Thus, the rate of the loss of the binary orbital angular momentum is estimated as 
\begin{equation}
 {\dot \mathcal{J}_{\rm{disk}}\over \mathcal{J}}
    \approx {|\dot{J}_{\rm{acc}}(r_{\rm{in}})|\over \mathcal{J}}
   = {\dot{M}r_{\rm{in}}^2
    \Omega_{\rm{in}}\over \mu a^2 \Omega_{\rm{orb}}}
   \approx \sqrt{2}{\dot{M}\over \mu}\,,
   \label{dotJaccoverJ}
\end{equation}
where $\Omega_{\rm{in}}$ is 
the Keplerian frequency of the circumbinary disk measured at 
$r_{\rm{in}}$ and $\dot{M}$ is set to be positive when the mass flux is inward.
In the last approximate equality 
$r_{\rm{in}}$ 
was approximated by $2a$~\citep{al94}.
Since currently we are not able to make the 
estimate of $\dot{\mathcal{J}}_{\rm{disk}}$ more precise, 
we introduce a parameter
\begin{equation}
j\equiv{\dot \mathcal{J}_{\rm{disk}}\over \mathcal{J}}  \left({\dot{M} \over \mu}\right)^{-1} \,,
\label{dotJoverJ}
\end{equation}
that represents the efficiency of angular momentum transfer 
from the binary to the circumbinary disk (recall that $\mu=\eta{M}$).

The angular momentum transfer from a circular equal-mass binary 
to the circumbinary disk has been recently studied 
by performing hydrodynamic simulations with $\alpha$-prescription \citep{mm08} 
and magneto-hydrodynamics (MHD) simulations \citep{sklh11}. 
In both simulations the sound velocity was set by hand instead of 
solving the energy equation. 
In such a numerical setup, one can freely scale 
the binary mass $M$, the semi-major axis $a$, and the surface density of the 
circumbinary disk $\Sigma$. 
\cite{mm08} derived the averaged mass accretion rate 
$\langle\dot{M}\rangle=2.5\times10^{-4}(GMa)^{1/2}\Sigma_0$ and the angular 
momentum flux 
carried by the circumbinary disk $\langle\dot{J}_{\rm{disk}}\rangle\simeq1.4
\times10^{-3}GMa\Sigma_0$ 
from the simulated result after $4000$ binary orbital cycles, 
where $\Sigma_0$ is a typical value of $\Sigma$ introduced for the normalization. 
On the other hand, \cite{sklh11} derived
$\langle\dot{M}\rangle=1.8\times10^{-2}(GMa)^{1/2}\Sigma_0$ and 
$\langle\dot{J}_{\rm{disk}}\rangle\simeq1.2\times10^{-2}GMa\Sigma_0$ 
after $\simeq 77$ binary orbital cycles.
Substituting these values of $\langle\dot{M}\rangle$ and $\langle\dot{J}_{\rm{
disk}}\rangle$ into equation~(\ref{dotJoverJ}), 
we can estimate the efficiency parameter $j$ to be $\sim 3.6$ for the former simulation 
and $\sim 0.69$ for the latter simulation, independent of 
$a$, $M$, and $\Sigma_0$. 
The scatter of $j$ between these two simulations 
will be mainly attributed to the difference in the strength of 
shear viscosity. 
\cite{mm08} assumed that the Shakura-Sunyaev viscosity parameter $\alpha_{\rm{SS}}$ 
to be $0.01$, whereas the effective value of $\alpha_{\rm{SS}}$ derived
from the MHD simulation by \cite{sklh11} was $\sim0.2$.
Namely, $j$ becomes larger for smaller viscosity. 
One can explain this tendency qualitatively as follows:  
when the mass accretion rate is fixed, 
the circumbinary disk becomes more massive for smaller viscosity, while 
a more massive disk gains larger resonant torque from the binary. 

The black hole growth due to the mass accretion 
from the circumbinary disk to each of the black holes via accretion disks 
can also modify the binary orbital evolution.
We note from equation~(\ref{dotJoverJ}) that the growth rate of binary mass 
$\dot{M}/M$ is the same order as the rate of the angular momentum transfer 
$\dot{\mathcal{J}}_{\rm{disk}}/\mathcal{J}$, if $j$ is the order of unity.
Hence, for simplicity, we 
absorb this mass growth effect into the coefficient $j$ in equation~(\ref{dotJoverJ}) 
and neglect the time variation of the binary mass in the following discussion. 

From the angular momentum conservation 
\begin{eqnarray}
\frac{d\mathcal{J}}{dt}
=
-(\dot \mathcal{J}_{\rm{gw}}+\dot \mathcal{J}_{\rm{disk}})\,,
\label{ecl}
\end{eqnarray}
we can derive the increasing rate of the GW frequency 
from equations~(\ref{binde}), (\ref{gwflux}), (\ref{dotJoverJ}), and (\ref{ecl})
as
\begin{eqnarray}
\dot{f}
&=&
\frac{96\pi}{5}
f^2
\left(\frac{\pi{G}\mathcal{M}f }{c^3}\right)^{5/3}
\nonumber \\
&&\times
\left[
1+
\frac{5}{32}
\frac{j}{\eta^{2}}
\frac{G\dot{M}}{c^3} 
x^{-4}+\cdots
\right].
\label{fdot}
\end{eqnarray}
The first term in the square brackets of equation~(\ref{fdot}) 
is the leading contribution from the quadrupole GW radiation given in 
equation~(\ref{quadrupoledotf}). 
The second term represents the effect of the circumbinary disk, 
and the ellipsis stands for the higher order PN corrections.
Recalling that $x\equiv \left(\pi G Mf/c^3\right)^{2/3}=O((v/c)^2)$,   
we note that the correction due to the binary-disk interaction 
has ``-4PN'' frequency dependence relative to the leading term for a 
constant $j$. 

By integrating equation~(\ref{fdot}) with respect to time,
we can express $f$ as a function of $t$, which in turn leads to
the expression of $t$ as a function of $f$:
\begin{eqnarray}
t(f)
&=&
t_0-\frac{5}{256\pi}\frac{1}{f}\left(\frac{\pi{G}\mathcal{M}f}{c^3}\right)^{-5/3}
\nonumber \\
&&\times
\left[1-
\frac{5}{64}\frac{j}{\eta^{2}}\frac{G\dot{M}}{c^3} 
x^{-4}+\cdots
\right].
\label{tgw2}
\end{eqnarray}
The GW phase defined by equation~(\ref{gwp}) then becomes
\begin{eqnarray}
\phi(f)
&=&
\phi_0-\frac{1}{16}\left(\frac{\pi G\mathcal{M}f}{c^3}\right)^{-5/3}
\nonumber \\
&&\times
\left[1-\frac{25}{416}
\frac{j}{\eta^{2}}
\frac{G\dot{M}}{c^3} 
x^{-4}+\cdots
\right].
\label{gwphase2}
\end{eqnarray}
From the Fourier transformation of equation~(\ref{ht}), we finally
find that the sky-averaged GW waveform in the Fourier domain
under the stationary phase approximation \citep{cf94, berti05} is given by
\begin{eqnarray}
\tilde{h}(f)=\frac{1}{2\sqrt{10}\pi^{2/3}}\frac{c}{D_L}\left(\frac{G\mathcal{M}}{c^3}\right)^{5/6}
f^{-7/6}e^{i\Psi(f)},
\label{hf}
\end{eqnarray}
where the sky-averaged GW phase:
\begin{eqnarray}
\Psi(f)
&=&
2\pi{f}t_0-\phi_0-\frac{\pi}{4}
+\frac{3}{128}\left(\frac{\pi{G}\mathcal{M}f}{c^3}\right)^{-5/3}
\nonumber \\
&&\times
\left[1-\frac{25}{832}
\frac{j}{\eta^{2}}
\frac{G\dot{M}}{c^3} 
x^{-4}
+
\cdots
\right].
\label{phase}
\end{eqnarray}
%
\subsection{Determination accuracy of mass accretion rates: rough estimate}
\label{subsec:daes1}
%
In this section, we analytically estimate the determination accuracy of 
the mass accretion rate of the circumbinary disk with {\it DECIGO/BBO}.
The determination accuracy of the mass accretion rate 
is mainly governed by the low frequency side in the observation band, 
where the effect of the circumbinary disk is the largest.
The lower cutoff frequency is determined by 
the duration of the observation or the detector sensitivity,
\begin{equation}
\fmin = \max(f(t=t_{\rm{ini}}),\flow),
\label{fmin}
\end{equation}
where $f(t=t_{\rm{ini}})$ 
is given by equation~(\ref{fobs}) into which we substitute 
an appropriate observation period $t_0-t_{\rm{ini}}$. 
We also adopt $\flow=10^{-3} \mathrm{Hz}$ 
for the cutoff due to the detector sensitivity of {\it DECIGO/BBO}.
Then, $\fmin$ turns out to be determined by $f(t=t_{\rm{ini}})$ 
for $t_0-t_{\rm{ini}}=5\,\rm{yr}$
in the mass range $6M_\odot<M\le5\times10^3M_\odot$. 

The square of the signal-to-noise ratio is defined by
\begin{equation}
SN^2
\equiv 
\int_{\fmin}^{\fmax}\varrho^2(f) d\ln{f},
\label{sn2}
\end{equation}
where 
\begin{equation}
\varrho^2(f)\equiv4N_\mathrm{dev}\frac{|\tilde{h}(f)|^2f}{S_n(f)}
\label{varrho}
\end{equation}
and
$N_\mathrm{dev}=8$ denotes the effective number of interferometers for 
{\it DECIGO/BBO} \citep{cd09} and $S_n(f)$ represents a total noise spectral density. 
As given in equation~(36) of \cite{yagi11}, we adopt 
the non sky-averaged instrumental noise spectral density for {\it DECIGO/BBO} as
\begin{eqnarray}
S^{\rm{inst}}_n(f) 
&=&
1.8\times 10^{-49}\,[\rm{Hz}^{-1}]\left(\frac{\it f}{1 \mathrm{Hz}} \right)^2 +2.9\times 10^{-49}\,[\rm{Hz^{-1}}]
\nonumber \\
&&+9.2\times 10^{-52}\,[\rm{Hz^{-1}}]\left(\frac{\it f}{1 \mathrm{Hz}} \right)^{-4} .
\label{noise-inst}
\end{eqnarray}  
This is obtained by multiplying the sky-averaged noise spectrum,  
which is shown in equation~(15) of \cite{cd09}, by 3/20 \citep{berti05}. 
The third term of the above equation dominates over the other terms at low frequencies so that $S^{\rm{inst}}_n(f)\approx9.2\times 10^{-52}\,[\rm{Hz^{-1}}]\left({\it f}/1 \mathrm{Hz} \right)^{-4}$. For simplicity, we assume in this subsection that 
\begin{equation}
S_n(f)=9.2\times10^{-52}\,[{\rm{Hz^{-1}}}]\left(\frac{f}{1\rm{Hz}}\right)^{-4}
\label{noise-neg}
\end{equation}
by neglecting the white dwarf confusion noise from the total noise spectral density. 
Note that the value of $\fmax$ is irrelevant in the following discussion of this subsection. 

To assess how accurately we can constrain the mass accretion rate, 
we extract the phase correction coming from the interaction between the 
binary and the circumbinary disk
in equation~(\ref{phase}), {\it i.e.,}
\begin{equation}
\Psi_{\mathrm{disk}}(f) = -\frac{3}{128} \frac{25}{832}
\frac{j}{\eta^{2/5}}
\frac{G\dot{M}}{c^3}
\left(\frac{\pi{G\mathcal{M}}f}{c^3}\right)^{-13/3}\,.
\label{sgwpdisk}
\end{equation}
If there are no degeneracies between the mass accretion rate and the other binary parameters, 
we can roughly claim that the 
effect of the circumbinary disk is detectable 
at the first order in the mass accretion rate as long as
the inequality 
\begin{equation}
|\varrho(f)\Psi_{\rm disk}(f)| \gtrsim 1
\label{rough}
\end{equation}
is satisfied. 
This inequality is most likely satisfied at $f=f_{\rm{min}}$ because $\Psi_{\rm{disk}}(f)$ is proportional to 
a large negative power of $f$. Therefore, we apply $f=f_{\rm{min}}$ to equation~(\ref{rough}).
As a typical source of {\it DECIGO/BBO}, 
we obtain $\varrho(f_{\rm{min}})\simeq108\,(D_{L}/3\,{\rm{Gpc}})^{-1}((t_0-t_{\rm{ini}})/5\,\rm{yr})^{-1/2}$ for five year observation.
Since $\varrho(f)$ is found to be proportional to ${M^{5/6}f^{4/3}}$ 
from equations~(\ref{hf}), (\ref{varrho}), and (\ref{noise-neg}), it exhibits no dependence on $M$ 
as long as we adopt $\fmin=f(t_{\rm{ini}})\propto M^{-5/8}$.
Then, the condition~(\ref{rough}) 
gives an analytic estimate 
for the determination accuracy of the mass accretion rate, $\Delta\dot{M}$, as 
\begin{equation}
\frac{\Delta\dot{M}}{\dot{M}_{\mathrm{Edd}}} 
\approx 
4.5\times10^{-4}\,
{\it j}^{-1}
\epsilon_{0.1}
\eta_{0.25}^{-13/8}
m_{10}^{5/8}
\left(\frac{t_0-t_{\rm{ini}}}{5\rm{yr}}\right)^{-13/8}
\label{mdot:low}
\end{equation}
in the mass range $6M_\odot\le{M}\lesssim3\times10^3M_\odot$, 
where we normalized the determination accuracy by the Eddington accretion rate:
\begin{eqnarray}
\dot{M}_{\rm{Edd}}=\frac{1}{\epsilon}\frac{L_{\rm{Edd}}}{c^2}\simeq
1.4\times10^{19}[\rm{g\,s^{-1}}]\,
\epsilon_{0.1}^{-1}
{\it m}_{10}\,,
\label{eddrate}
\end{eqnarray}
and $L_{\rm{Edd}}=4\pi GMm_{\rm{p}}c/\sigma_{\rm{T}}$ 
is the Eddington luminosity with $m_{\rm{p}}$ 
and $\sigma_{\rm{T}}$ denoting the proton mass and 
Thomson scattering cross section, respectively. 
The mass-to-energy conversion efficiency $\epsilon$ 
is set to $0.1$ in the following discussion.

The relative accuracy in comparison with the Eddington rate
decreases with the total mass of the binary but increases in proportion to $j$.
This can be understood by the facts that the Eddington rate is proportional
to $M$ and that 
$\Psi_{\rm{disk}}(f_{\rm min})\propto{\it j}{M}^{-13/8}$
in equation~(\ref{sgwpdisk}), which gives $\Delta\dot{M}\propto {\it j}^{-1}M^{13/8}$. 
We will compare this analytic estimate with a more precise numerical 
one in the succeeding subsection.

%
\subsection{Determination accuracy of mass accretion rates: more precise estimate}
\label{subsec:daes2}
%

We derive the determination accuracy of the mass accretion rate 
by applying the matched filtering analysis \citep{cf94}. 
We adopt the phase correction given in equation~(\ref{phase})
as templates, although we do not know the exact form of 
the correction to the GW phase.  
More detailed study for more reliable prediction of 
the phase correction is left for future work. 
By neglecting the black hole spins, the determination accuracy of the binary parameters 
$\theta^a=\{\ln{\mathcal{M}}, \ln\eta, t_0, \phi_0, D_L, \dot{M}\}$ 
can be estimated by $\Delta\theta^a =\sqrt{\left(\Gamma^{-1}\right)_{aa}/N_{\rm{dev}}}$, 
where 
\begin{equation}
\Gamma_{ab} 
\equiv 
4 \mathrm{Re} 
\int_{f_{\mathrm{min}}}^{f_{\mathrm{max}}} 
\frac{\partial\tilde{h}}{\partial \theta^a} 
\frac{\partial\tilde{h}^*}{\partial \theta^b} 
\frac{1}{S_n(f)} df
\end{equation}
is the Fisher matrix. 
Here, 
\begin{equation}
\fmax = \min(\fisco,\fhigh)
\label{fminmax}
\end{equation}
is the cutoff frequency on the higher frequency side, where 
the cutoff frequency determined by the detector sensitivity of {\it DECIGO/BBO} is taken as 
$\fhigh = 100 \mathrm{Hz}$ and 
$\fisco$ is the GW frequency at the ISCO given in equation~(\ref{fisco}).
For the noise spectrum $S_n(f)$,
we adopt the total noise spectral density (see equation~(36) of \cite{yagi11}), 
which includes the confusion noise of WD/WD binaries that masks the instrumental noise below $f<0.2$\,Hz \citep{fp03}. 
We also assumed that the NS/NS foreground noise can be cleaned 
down to the level below the instrumental noise.

Figure~\ref{fig:da} shows to what extent one can constrain 
the mass accretion rate of coalescing binary black holes 
from the GW observation with {\it DECIGO/BBO}.
The vertical axis shows the standard error of the mass accretion rate 
normalized by the Eddington rate,
whereas the horizontal axis shows the total mass of binary black holes.
The mass accretion rate is measurable at a statistically 
significant level, if it is greater than the determination accuracy $\Delta\dot{M}$. 

The solid, dashed, and dotted curves represent $\Delta\dot{M}/\dot{M}_{\rm{Edd}}$ 
for the cases of the observation time of $t_0-t_{\rm{ini}}=$1, 3, and 5\,yr, respectively.
The condition $\Delta{\dot{M}}\le\dot{M}_{\rm{Edd}}$ is realized 
in the mass range $6M_\odot\le{M}\lesssim3\times10^3M_\odot$ for one year observation 
and in the mass range $6M_\odot\le{M}\lesssim8\times10^3M_\odot$ 
for both three year and five year observations.
The determination accuracy is higher for binaries at a closer distance 
from the earth and is also higher with a longer observation time. 
We also note that the dependence of the determination accuracy 
on the observation time is in good agreement with $(t_0-t_{\rm{ini}})^{-13/8}$ shown in equation~(\ref{mdot:low}) 
in the mass range $50M_\odot\lesssim{M}\lesssim3\times10^{3}M_\odot$.
From Figure~\ref{fig:da}, we can estimate the mass dependence of 
the determination accuracy in five year observation as
\begin{equation}
\frac{\Delta\dot{M}}{\dot{M}_{\mathrm{Edd}}} 
\approx 
1.0\times10^{-2}\,
j^{-1}
m_{10}^{10/23}
\label{mdot:fig}
\end{equation}
in the mass range $6M_\odot\lesssim{M}\lesssim3\times10^3M_\odot$. 
By comparing this estimate with equation~(\ref{mdot:low}), 
we find that the analytic estimate tends to give about one order of 
magnitude higher determination accuracy and 
that its dependence on the binary mass is slightly different. 
These are because we have neglected the contribution 
from the white dwarf confusion noise in the analytic estimate.

%
\section{Expected number of merger events}
\label{sec:nmevents}
%

In this section, we estimate 
the expected number of merger events that have detectable signatures 
of the binary-disk interaction in GWs. 
\subsection{Circumbinary disk formation}
\label{subsec:cbdisk}
%
The presence of a circumbinary disk around binary black holes with 
sufficiently large accretion rate will not be ubiquitous. 
To achieve such a large accretion rate as shown in equation~(\ref{mdot:fig}), 
it will be required that binary black holes are embedded in a dense gaseous 
environment like a molecular cloud which exists in a star forming region of a galaxy. 
Its typical length scale of the molecular cloud is $10^{2-3}\,\rm{pc}$ 
and its molecular hydrogen number density is distributed 
over the range of $10^{3-6}\,\rm{cm^{-3}}$.

If we consider a binary traveling through a dense molecular cloud, 
the circumbinary disk will be formed around the binary 
through the Bondi-Hoyle-Lyttleton (BHL) accretion. 
The gas within the BHL radius expressed by
\begin{equation}
r_{\rm{BHL}}
=
\frac{2GM}{{\it v}_\infty^2}
\sim
2.2\times10^{-4}[\rm{pc}]\,
{\it v}_{20}^{-2}
{\it m}_{10}\,
\label{rbhl}
\end{equation}
will accrete onto the binary, 
where $v_{20}\equiv{v}_\infty/(20\,\rm{km\,s^{-1}})$ and $v_\infty\equiv\sqrt{v_{\rm{bulk}}^2+c_{\rm{s},\infty}^{2}}$ 
with the sound velocity $c_{\rm{s},\infty}=0.2\,\rm{km\,s^{-1}}$ for typical molecular clouds. 
Here, we also adopted $v_{\rm{bulk}}=20\,\rm{km\,s^{-1}}$ as 
a fiducial bulk velocity of the binary relative to the ambient medium. 
This is indicated from the observations that the radial velocities 
of some X-ray binaries including GRO\,J0422+32 in our galaxy 
are $\sim10\,\rm{km\,s^{-1}}$ (see Table~1 of \citet{nv99}).
Since the BHL radius is much less than a typical size of a molecular cloud
core $\sim 1\,\rm{pc}$, the circumbinary disk can be 
formed even in the core of a molecular cloud.

While the outer edge radius of the circumbinary disk should be less than the BHL radius, 
its inner edge radius is approximately given by $r_{\rm{in}}\approx2a(t)$ \citep{al94}.
The consistency condition $r_{\rm{BHL}}>r_{\rm{in}}$ requires 
$a(t)<a_{\rm{BHL}}
\sim1.1\times10^{-5}\,[\rm{pc}]\,
{\it v}_{20}^{-2}m_{10}$, where $a(t)$ can be read from equation~(\ref{acoales}). 
At five years before the coalescence, this condition is easily satisfied. 

The rate of the BHL accretion is given by \citep{bh44}
\begin{eqnarray}
\dot{{\it M}}_{\rm{BHL}}
&=&
4\pi(G{\it{M}})^{2}\rho_\infty v_\infty^{-3}
\nonumber \\
&\sim&
2.8\times10^{17}\,[\rm{g\,s^{-1}}]\,
\rho_{-19}
{\it v}_{20}^{-3}
{\it m}_{10}^2\,,
\label{bhlrate}
\end{eqnarray}
where $\rho_{-19}\equiv\rho_\infty/(10^{-19}\,\rm{g\,cm^{-3}})$ 
with the density of the molecular cloud $\rho_\infty$. 
Its fiducial value $10^{-19}\,\rm{g\,cm^{-3}}$ corresponds 
to a typical density of dense molecular cloud cores.
In Figure~1, the dash-dotted line represents this BHL rate 
normalized by the Eddington rate. 
The determination accuracy is higher than $\dot{M}_{\rm{BHL}}$ in the mass range 
$M\gtrsim100M_\odot$ for  one year observation and in the almost whole mass range 
for both three year and five year observations.
It can happen that the mass accretion rate is in the range 
between $\Delta\dot{M}$ and $\dot{M}_{\rm{BHL}}$, 
and then the effect of the circumbinary disk is 
detectable at a statistically significant level with {\it DECIGO/BBO}.

From equations~(\ref{eddrate}) and (\ref{bhlrate}), 
the lowest number density of the molecular cloud for a BHL rate to be 
measured with the determination accuracy $\Delta{\dot{M}}$
is given by
\begin{equation}
n_{\rm{MC}}\sim1.5\times10^6\,[\rm{cm^{-3}}]\,
\epsilon_{0.1}^{-1}
{\it v}_{20}^{3}
{\it m}_{10}^{-1}
\left(\frac{\Delta\dot{\it M}}{\dot{\it M}_{\rm{Edd}}}\right),
\label{nmc}
\end{equation}
where we substitute $\Delta\dot{M}$ into $\dot{M}_{\rm{BHL}}$.

%
\subsection{Estimate of merger events}
\label{subsec:events}
%
Following \cite{mt05}, 
the volume fraction of molecular clouds with $n\gtrsim{n_{\rm{MC}}}$ in
our galaxy can be modeled as
\begin{eqnarray}
f_{\rm{MC}}
&=&
\frac{(\beta-2)\langle\Sigma_{\rm{MC}}\rangle}{2\mu{m_{\rm{p}}}n_{\rm{min}}^2H_{\rm{g}}}
\int_{n_{\rm{MC}}}^{n_{\rm{max}}}\left(\frac{n}{n_{\rm{min}}}\right)^{-\beta}dn
\nonumber \\
&\sim&
5.2\times10^{-10}
\epsilon_{0.1}^{9/5}
v_{20}^{-27/5}
m_{10}^{9/5}
\left(\frac{\Delta\dot{M}}{\dot{M}_{\rm{Edd}}}\right)^{-9/5},
\label{fmc}
\end{eqnarray}
where $n$ is the molecular hydrogen number density, 
$n_{\rm{min}}=10^2\rm{cm^{-3}}$ 
and $n_{\rm{max}}=10^5\rm{cm^{-3}}$ are the minimum and maximum number densities
of typical molecular clouds, $\beta$ is the index of power-law distribution,
$\langle\Sigma_{\rm{MC}}\rangle$ is the mean surface mass density of molecular clouds,
and $H_{\rm{g}}$ is the scale-height of the Galactic disk.
Here, we adopted the values of $\beta=2.8$ \citep{ak02},
$\langle\Sigma_{\rm{MC}}\rangle=29\,M_\odot/\rm{pc}^2$, 
and $H_{\rm{g}}=75\,\rm{pc}$ \citep{sa84}.
From the condition that $n_{\rm{max}}$ should not exceed $n_{\rm{MC}}$, 
we obtain the upper limit of $\Delta\dot{M}/\dot{M}_{\rm{Edd}}$ as
\begin{equation}
\frac{\Delta\dot{M}}{\dot{M}_{\rm{Edd}}}
\lesssim
6.7\times10^{-2}
\epsilon_{0.1}
m_{10}
v_{20}^{-3}.
\label{mdotmax}
\end{equation}

The merger rate of binary stellar-mass black holes due to the GW emission 
in our galaxy is given by $f_{\rm{BH}}=2.56\times10^{-5}\,\rm{yr^{-1}}$ 
by ``Model A'' of Table~4 of \cite{bel02}.
The expected number of merger events for the binary black holes within $3\,\rm{Gpc}$ 
for five year observation is then estimated as
\begin{eqnarray}
N
&\sim&
\frac{4\pi}{3}
n_{\rm{gal}} 
f_{\rm{BH}}
f_{\rm{MC}}
\left(\frac{D_{L}}{1+z}\right)^3
\frac{t_0-t_{\rm{ini}}}{1+z}
\nonumber \\
&\sim&
6.0\times10^{-2}\,
j^{9/5}
\epsilon_{0.1}^{9/5}
v_{20}^{-27/5}
n_{\rm{gal},0.01}
m_{10}^{117/115},
\label{mgrate}
\end{eqnarray}
where $n_{\rm{gal},0.01}\equiv{n_{\rm{gal}}}/(0.01\,\rm{Mpc^{-3}})$ with 
the number density of galaxies $n_{\rm{gal}}$ and 
the redshift $z$ is set to $0.5$ corresponding to 
$D_{L}=3\,\rm{Gpc}$. We also estimate $f_{\rm{MC}}$ in the above equation 
by substituting equation~(\ref{mdot:fig}) into equation~(\ref{fmc}). 
Note that $\Delta\dot{M}/\dot{M}_{\rm{Edd}}$ satisfies the condition given by equation~(\ref{mdotmax}).
We have not taken into account the redshift evolution 
of the merging rate of the binaries per galaxy in the above analysis. 
When this effect is added, the number of events is expected to be a few times larger. 
We conclude that the feasibility of detection of circumbinary disks around 
binary stellar-mass black holes 
is marginal in five year observation with ${\it DECIGO/BBO}$. 
We emphasize that our analysis so far does not rely on specific models 
of the circumbinary disk structure.

%
\section{Astrophysical implications of binary black holes with circumbinary disks}
%

In this section, we further discuss some astrophysical implications of binary 
black holes with a circumbinary disk.

%
\subsection{Evolution of binary black holes in the molecular cloud}
%

The binary transfers its orbital energy and angular momentum 
to the circumbinary disk through the tidal/resonant interaction, 
which leads to the rapid orbital decay of the binary.
If the binary evolution towards the coalescence is accelerated
significantly during its stay in the molecular cloud, 
the merger rate of binary black holes would be increased 
in comparison with the previous estimates of \cite{bel02}.
In this subsection, we examine this possibility.

Let us consider a circular binary whose coalescence timescale 
due to the GW emission given by \citep{p64}
\begin{eqnarray}
t_{\rm{gw}}
&=&
\frac{5}{64}
\left(\frac{a}{{\it r}_{\rm{g}}}\right)^4
\frac{r_{\rm{g}}}{c}
\frac{1}{\eta}
\nonumber \\
&\sim&
1.0\times10^{-12}\,[\rm{yr}]\,
{\it m}_{10}
\eta_{0.25}^{-1}
\left(\frac{\it a}{{\it r}_{\rm{g}}}\right)^4
\label{tgw}
\end{eqnarray}
exceeds the Hubble time.
The binary-disk interaction will be one of the key mechanisms to force such binary 
to merge within the Hubble time. The orbital decay timescale due to the tidal/resonant interaction 
corresponds to the inverse transfer rate of the orbital angular momentum, 
which is given by equation~(\ref{dotJoverJ}) as
\begin{eqnarray}
t_{\rm{tide/res}}
=\frac{1}{j}\frac{\mu}{\dot{M}}\sim
1.1\times10^{7}\,[\rm{yr}]\,
{\it j}^{-1}
\epsilon_{0.1}
\eta_{0.25}
\dot{\it m}^{-1}.
\end{eqnarray}
This is consistent with equation~(17) of \cite{hui10} when $j=1.0$ is adopted.
Owing to this tidal/resonant interaction, the orbital evolution is accelerated 
even when $t_{\rm{gw}}>t_{\rm{tide/res}}$ holds, {\it i.e}., 
in the range $a_{\rm{tide/res}}<a\le{a_{\rm{BHL}}}$ with
\begin{eqnarray}
\frac{a_{\rm{tide/res}}}{r_{\rm{g}}}
&=&
\left(\frac{64}{5}
\frac{ct_{\rm{tide/res}}}{r_{\rm{g}}}
\eta
\right)^{1/4}
\nonumber \\
&\sim&
6.9\times10^4\,
{\it j}^{-1/4}
\epsilon_{0.1}^{1/4}
\eta_{0.25}^{1/2}
m_{10}^{-1/4}
\dot{m}^{-1/4},
\label{ac}
\end{eqnarray}
if the binary stays in the molecular cloud for a sufficiently long term.

Since the typical size of a giant molecular cloud is 
${\it r}_{\rm{MC}}\sim100\,\rm{pc}$, the crossing time of binary black holes 
in the molecular cloud is $t_{\rm{cross}}\sim5\times10^6\,[\rm{yr}]\,{\it r}_{\rm{MC,100}}
({\it v}_{\rm{bulk}}/20\rm{km\,s^{-1}})^{-1}$, where $r_{\rm{MC},100}\equiv r_{\rm{MC}}/(100\,\rm{pc})$. 
During this crossing time, the binary can keep 
interacting with the molecular cloud gas.
The condition $t_{\rm{cross}}\ge t_{\rm{tide/res}}$ leads to $\dot{M}\ge\dot{M}_{\rm{c}}$ 
with the critical accretion rate 
\begin{eqnarray}
\dot{M}_{\rm{c}}
\sim
2.3\,
{\it j}^{-1}
v_{\rm{20}}
\epsilon_{0.1}
\eta_{0.25}
r_{\rm{MC},100}^{-1}\,
\dot{M}_{\rm{Edd}}.
\end{eqnarray}
If $\dot{M}\ge\dot{M}_{\rm{c}}$,
binary black holes can merge 
within the Hubble time, even if the initial separation 
is too large to coalesce within the Hubble time by means of the radiation reaction 
due to the GW emission alone.

%
\subsection{Inner-most disk mass}
%

In this subsection, 
we evaluate the mass of the circumbinary disk with the aid of the assumption that 
the structure of the circumbinary disk is described by the {\it standard disk model} \citep{ss73}.
The circumbinary disk is assumed to be a steady, axisymmetric, 
geometrically thin, gas-pressure and electron-scattering opacity dominated, Keplerian disk. 
Here, we use the following solution of the standard disk \citep{kato08}:
\begin{eqnarray}
\Sigma(r)
&\approx&
4.1\times10^5\,[\rm{g\,cm^2}]\,
\nonumber \\
&&\times
\alpha_{\rm{SS}}^{-4/5}
\epsilon_{0.1}^{-3/5}
m_{10}^{1/5}
\dot{m}^{3/5}
\left(\frac{\it r}{{\it r}_{\rm{g}}}\right)^{-3/5},
\label{sig}
\\
H(r)
&\approx&
1.6\times10^{4}\,[\rm{cm}]\,
\nonumber \\
&&\times
\alpha_{\rm{SS}}^{-1/10}
\epsilon_{0.1}^{-1/5}
m_{10}^{9/10}
\dot{m}^{1/5}
\left(\frac{\it r}{{\it r}_{\rm{g}}}\right)^{21/20},
\label{hr}
\\
c_{\rm{s}}(r)
&\approx&
1.7\times10^{8}\,[\rm{cm\,s^{-1}}]\,
\nonumber \\
&&\times
\alpha_{\rm{SS}}^{-1/10}
\epsilon_{0.1}^{-1/5}
m_{10}^{-1/10}
\dot{m}^{1/5}
\left(\frac{\it r}{{\it r}_{\rm{g}}}\right)^{-9/20},
\label{cs}
\end{eqnarray}
where $\Sigma$, $H$, $c_{\rm{s}}$, 
$r_{\rm{g}}\equiv{GM}/c^2$, and $\dot{m}\equiv\dot{\it M}/\dot{\it M}_{\rm{Edd}}$ 
are the surface density, the scale-height of the disk, 
the sound velocity measured at the mid-plane temperature of the disk, 
the gravitational radius, and the normalized mass accretion rate, 
respectively (recall that $\alpha_{\rm{SS}}$ is the Shakura-Sunyaev viscosity parameter).

As it is not the whole mass of the circumbinary disk 
but the mass at around the inner edge of the circumbinary disk 
that affects the GW phase evolution, 
we estimate 
\begin{eqnarray}
M_{\rm{d,in}}=\pi r_{\rm{in}}^2\Sigma(r_{\rm{in}})
\label{eq:dm}
\end{eqnarray}
and refers to it as an inner-most disk mass. 
It can be evaluated by substituting $r_{\rm{in}}=2a$ 
into the expression of $\Sigma$ given in equation~(\ref{sig}).

Figure~\ref{fig:dm} shows the inner-most disk mass of the circumbinary
disk that is detectable by GWs. They are evaluated at the beginning of the 
observation period of $t_0-t_{\rm{ini}}=$$1$, $3$, and $5$ yr, respectively. 
We also adopt $\alpha_{\rm{SS}}=0.1$.
It shows that the existence of the circumbinary disk can be confirmed,
even if the disk mass is less than the binary mass by many orders of 
magnitude.
This is because we can observe many ($N_{\rm cyc} \sim 10^6$) cycles 
of GWs.

We can understand the binary mass dependence of $M_{\rm{d,in}}/M$ 
in Figure~\ref{fig:dm} in the following way.
The semi-major axis for a fixed value of $t_0-t_{\rm{ini}}$ 
is proportional to $M^{3/4}$ from equation~(\ref{acoales}) 
and the surface density at the inner edge of the circumbinary disk is proportional to 
$M^{1/5}\dot{M}^{3/5}a^{-3/5}$ from equation~(\ref{sig}).
When $\dot{M}$ is given in equation~(\ref{mdot:low}), we find
\begin{equation}
\frac{M_{\rm{d,in}}}{M} 
\approx
2.2 \times 10^{-12}
m_{10}^{49/40}
\end{equation}
in the mass range $6M_\odot\le{M}\lesssim3\times10^3M_\odot$ for five year observation. 
This is roughly in agreement with the dotted curve in Figure~\ref{fig:dm}.

%
\subsection{Decoupling from the circumbinary disk}
\label{Sec:decoupling}
%

When $t_{\rm{gw}}$ of the binary is shorter than the viscous timescale evaluated 
at the inner edge of the circumbinary disk, the binary is decoupled from the circumbinary disk.
After the decoupling, no binary-disk interaction occurs so that 
it is impossible to detect the effect of the circumbinary disk by the GW observation.

The viscous timescale measured at the inner edge of the circumbinary disk 
is given by
\begin{eqnarray}
t_{\rm{vis}}(a)
=
\frac{r_{\rm{in}}^2}{\nu}
\approx 
t_{\rm{vis,0}}\left(\frac{a}{r_{\rm{g}}}\right)^{7/5},
\label{tvis}
\end{eqnarray}
where we used the relation for the disk viscosity $\nu=\alpha_{\rm{SS}}c_{\rm{s}}H$ 
and $t_{\rm{vis,0}}\equiv t_{\rm{vis}}(r_{\rm{g}})\approx(1/\alpha_{\rm{SS}})(cr_{\rm{g}}/c_{\rm{s}}^2(r_{\rm{g}}))$ 
with equations~(\ref{hr}) and (\ref{cs}).
The decoupling radius at which $t_{\rm{gw}}(a)=t_{\rm{vis}}(a)$ is given by
\begin{eqnarray}
\frac{a_{\rm{d}}}{r_{\rm{g}}}
&=&
\left(
\frac{64}{5}
\frac{ct_{\rm{vis,0}}}{r_{\rm{g}}}\eta
\right)^{5/13}
\nonumber \\
&\sim&
1.1\times10^2\,
\alpha_{\rm{SS}}^{-4/13}
\eta_{0.25}^{5/13}
\epsilon_{0.1}^{2/13}
m_{10}^{1/13}
\dot{m}^{-2/13}.
\label{ad}
\end{eqnarray}
Even when $t_{\rm{gw}}$ is shorter than $t_{\rm{tide/res}}$, 
the binary continues to gravitationally interact with the circumbinary disk 
until the binary orbit decays down to $a_{\rm{d}}$.
In the mass range of $6M_\odot\le{M}\lesssim2\times10^4M_\odot$,
the condition $a\gtrsim {\it a}_{\rm{d}}$ is satisfied 
even at one year before the coalescence. 
Namely, the binary continues to 
gravitationally interact with the circumbinary disk, as we described, 
at least at the early stage of the GW phase evolution 
during the observation period.
Thus, although we take no account of 
the decoupling effect in Figure~\ref{fig:da}, 
this neglect has been justified 
in the mass range $M\lesssim10^3M_{\odot}$. 
For the larger mass range, 
the determination accuracy is deteriorated 
by taking into account the decoupling effect. 
However, for $M>1.4 \times 10^4 M_{\odot}$, 
{\it DECIGO/BBO} is completely insensitive to the circumbinary disk 
when the mass accretion rate is smaller than the Eddington rate.

The BHL accretion rate can exceed the Eddington rate 
depending on parameters: $\rho_{\infty}$, $v_{\infty}$, and $M$. 
X-ray observations have detected the bright X-ray sources, 
such as GRS~1915+105 in our galaxy, with a luminosity
over the Eddington one (cf. \citealt{done07}). 
Such large luminosities can be explained by a supercritical 
accretion flow onto stellar-mass black holes. 
If the circumbinary disk is in a super-critical state, equation~(\ref{ad}) does not apply.
Since the accretion timescale is much shorter than the viscous timescale in the super-critical state,
the binary will not be decoupled from the circumbinary disk until the coalescence,
although it is still unclear how such a circumbinary disk evolves \citep{tk10}.

%
\section{Summary and discussion}
%

We have derived the minimum mass accretion rate of a circumbinary disk 
around binary stellar/intermediate-mass black holes 
necessary for the detection by future space-borne GW detectors 
such as {\it DECIGO/BBO}.
Our main conclusions are summarized as follows:
\begin{enumerate}
\renewcommand{\theenumi}{\arabic{enumi}}
\item 
The circumbinary disk with 
$\dot{\it M}\sim1.4\times10^{17}[\rm{g\,s^{-1}}]\,{\it j}^{-1}
({\it M}/10{\it M}_\odot)^{33/23}$ with $10\%$ mass-to-energy conversion efficiency, 
where $j$ represents the efficiency of the angular momentum transfer defined by 
equation~(\ref{dotJoverJ}), within $3\,\rm{Gpc}$ 
from the earth is detectable at a statistically significant level 
with {\it DECIGO/BBO} by observing for five years before the coalescence. 
The determination accuracy of the mass accretion rate 
is higher at a closer distance from the earth, 
and is also higher for a longer observation time.
\item
If binary stellar-mass black holes are residing in 
sufficiently dense molecular clouds, 
the influence of the circumbinary disk will be detectable.   
The number of merger events of such binaries 
within $3\,\rm{Gpc}$ for five year observation, 
is approximately estimated to be $\mathcal{O}(0.1)$
by assuming typical values for the event rate of the black hole coalescence 
and that molecular clouds have the number density from $10^{-5}\,\rm{cm^{-3}}$
to $n_{\rm{MC}}$ given in equation~(\ref{nmc}).
This suggests that there might be a possibility to detect
the circumbinary disk around binary stellar-mass black holes 
with {\it DECIGO/BBO}.
\end{enumerate}

In the estimate presented in this paper, 
we still have a lot of uncertainties, {\it e.g.,} the formation 
rate of binary black holes, the molecular hydrogen number density, 
the relative velocity to the molecular cloud, the efficiency of angular momentum 
transfer (${\it j}$: see equation~(\ref{dotJoverJ})), and so on. 
Therefore, we should understand that there remains a possibility 
that the event rate is one or two orders of magnitude larger than our estimate. 

For the idealized situation that we discussed in this paper, 
the correction to the GW phase evolution shows $-4$PN frequency dependence.  
This dependence is the same as the one for a braneworld model discussed in \citet{yagi11}.
Similarly, the cosmic acceleration \citep{seto01} and the alternative theories of gravity 
in which the gravitational constant $G$ is not a costant \citep{yunesG} also give the corrections 
with $-4$PN frequency dependence.
Furthermore, the acceleration acting on a binary in time-independent external gravitational 
field gives the correction with the same frequency dependence \citep{yunes-perturber}. 
Therefore, once the 
deviation from the standard template is discovered, discriminating 
the effect of the circumbinary disk from the others will be a difficult task. 
However, 
for example, the binary-disk interaction rapidly 
decays after the decoupling of the circumbinary disk as we 
discussed in section~\ref{Sec:decoupling}.
This is a signature unique to the interaction with the circumbinary disk. 
In addition, there can be a method to distinguish the effect of the circumbinary disk from that of modified gravity theory
by detecting the GW signals from multiple sources. This is because the effect of modified gravity theory is universal, 
whereas that of circumbinary disk is not.
Therefore, further detailed studies may provide some ways of distinctions.

In this paper we have assumed that the binary is in a circular orbit.
The orbital eccentricity also gives the frequency dependence of 
negative PN order (-19/6PN) \citep{ch06}. 
Inclusion of the orbital eccentricity as a parameter of the waveform 
reduces the determination accuracy of $-4$PN order term  
by some factors.
However, we expect that we can determine both the orbital eccentricity and the mass accretion rate
for the large mass parameter range with errors of $\mathcal{O}(0.1)$ for typical binary parameters. 
\cite{kc11} calculated the determination accuracies of full seventeen
binary parameters including an initial orbital eccentricity for binary massive black holes 
with {\it LISA}, using a Bayesian analysis under Markov Chain Monte Carlo simulations.
They conclude that the {\it LISA} could distinguish between circular orbit and 
eccentric orbit with a very small orbital eccentricity $e\sim10^{-3}$.

Furthermore, we would like to point out that
the effect of the circumbinary disk can be important, 
even if it is suppressed to the undetectable level for each event. 
The primary mission of the space-borne GW antennas such as 
{\it DECIGO/BBO} is to measure the primordial GW background radiation. 
In order to achieve this goal, abundant binary GW sources become the foreground noises. 
To remove them, the expected GW signals for the events identified by the matched
filtering will be subtracted from the data.
However, if the theoretical templates contain 
some unknown elements, the subtraction leaves residuals 
that remain to contribute to confusion noises \citep{ch06,harmsetal08}. 

Finally, we briefly mention the relevance of our discussion 
to the ground-based GW detectors. 
With these detectors, the distance to which we can detect 
GWs from binary black holes is reduced, and hence the expected 
event rate is extremely small.
Using the third-generation-ground-based GW detector: 
Einstein Telescope (ET), and assuming high signal-to-noise ratio 
corresponding to an accidentally nearby event at $D_L=100$Mpc, 
the effect of the circumbinary 
disk will be detectable only for the mass accretion rate $10^4$ times 
higher than the Eddington rate.
Even if stellar-mass binary black holes are at rest in a dense molecular 
cloud core, the expected BHL accretion rate is $10^{3-4}$ 
times as much as the Eddington rate.
Hence, we would be able to safely conclude that the effect of the 
circumbinary disk becomes relevant only for space-borne 
GW detectors such as {\it DECIGO/BBO}.

%
\acknowledgments
%

The authors are grateful to the members of the regular meeting 
on gravity and gravitational waves (GG seminar) at 
Department of Physics, Kyoto University
for helpful and continuous discussions.
This work has been supported by the Grants-in-Aid 
of the Ministry of Education, Science, Culture, and Sport and Technology 
(MEXT;  21540304, 22540243, 23540271 KH, 22340045 SM, 21244033, 21111006, 22111507
TT) 
and in part by the Grant-in-Aid for the Global COE Program 
"The Next Generation of Physics, Spun from Universality and Emergence". 
K.Y.~is also supported by the Japan Society for the Promotion 
of Science grant No.~$22 \cdot 900$.



\begin{thebibliography}{}
%
\bibitem[Agol \& Kamionkowski(2002)]{ak02}
Agol,~E., \& Kamionkowski,~M. 2002, \mnras, 334, 553
%
\bibitem[Armitage \& Natarajan (2002)]{armi1}
Armitage,~P. J., \& Natarajan,~P. 2002, \apj, 567, L9
%
\bibitem[Artymowicz \& Lubow(1994)]{al94}
Artymowicz,~P., \& Lubow,~S.H. 1994, \apj, 421, 651
%
\bibitem[Artymowicz \& Lubow(1996)]{al96}
Artymowicz,~P., \& Lubow,~S.H. 1996, \apj, 467, L77
%
\bibitem[Barausse \& Rezzolla(2008)]{br08}
Barausse,~E., \& Rezzolla,~L. 2008, Phys.\ Rev.\ D, 77, 104027
%
\bibitem[Belczynski et al.(2002)]{bel02}
Belczynski,~K., Kalogera,~V., \& Bulik,~T. 2002, \apj, 572, 407
%
\bibitem[Berti et al.(2005)]{berti05}
Berti,~E., Buonanno,~A., \& Will,~C.M. 2005, Phys.\ Rev.\  D 71, 084025
%
\bibitem[Bode et al.(2010)]{bode10}
Bode,~T., Haas,~R., Bogdanovi\'c,~T., Laguna,~P., \& Shomarker,~D. 2010, \apj, 715, 1117
%
\bibitem[Bondi \& Hoyle(1944)]{bh44}
Bondi,~H., \& Hoyle,~F. 1944, \mnras, 104, 273
%
\bibitem[Bogdanovi\'c et al.(2008)]{bo08}
Bogdanovi\'c,~T., Smith,~B.D., Sigurdsson,~S., \& Eracleous,~M. 2008, \apj, 174, 455
%
\bibitem[Chang et al.(2009)]{cp09}
Chang,~P., Strubble,~L.E., Menou,~K., \& Quataret,~E. 2010, \mnras, 407, 2007 
%
\bibitem[Cuadra et al.(2009)]{cu09}
Cuadra, J., Armitage, P.J., Alexander, R.D., \& Begelman, M.C. 2009, \mnras, 393, 1423
%
\bibitem[Cutler \& Holz(2009)]{cd09}
Cutler,~C., \& Holz,~D.E. 2009, Phys.\ Rev.\ D, 80, 104009
%
\bibitem[Culter \& Flanagan(1994)]{cf94}
Culter,~C., \& Flanagan~E.E. 1994, Phys.\ Rev.\ D, 49, 2658
%
\bibitem[Cutler \& Harms(2006)]{ch06}
Cutler,~C., \& Harms,~J. 2006, Phys.\ Rev.\ D, 73, 042001
%
\bibitem[Danzmann(1997)]{d97}
Danzmann,~K. 1997, Class.\ Quant.\ Grav., 14, 1399
%
\bibitem[Done et al.(2007)]{done07}
Done, C., Gierlinski, M., \& Kubota, A. 2007, A\&AR, 15, 1 
%
\bibitem[Harms et al.(2008)]{harmsetal08}
Harms,~J., Christoph,~M., Markus,~O., \& Malte,~P. 2008, Phys.\ Rev.\ D, 77, 123010
%
\bibitem[Hayasaki et al.(2008)]{hmh08}
Hayasaki,~K., Mineshige,~S., \& Ho,~C.L. 2008, \apj, 682, 1134
%
\bibitem[Hayasaki et al.(2007)]{hms07}
Hayasaki,~K., Mineshige,~S., \& Sudou,~H. 2007, \pasj, 59, 427
%
\bibitem[Hayasaki(2009)]{kh09}
Hayasaki, K. 2009, \pasj, 61, 65
%
\bibitem[Hayasaki et al.(2010)]{hui10}
Hayasaki,~K., Ueda,~Y., \& Isobe,~N. 2010, \pasj, 62, 1351
%
\bibitem[Holz \& Hughes(2005)]{hh05}
Holz,~E.D., \& Hughes,~A.S. 2005, \apj, 629, 15
%
\bibitem[Kato et al.(2008)]{kato08}
Kato,~S., Fukue,~J., \& Mineshige,~S. {\it Black-Hole Accretion Disks}, (Kyoto Univ. Press. 2008).
%
\bibitem[Key \& Cornish(2011)]{kc11}
Key,~J.S., \& Cornish,~N.J. 2011, Phys.\ Rev.\ D, 83, 3001
%
\bibitem[Kocsis et al.(2011)]{kocsis11}
Kocsis,~B., Yunes,~N., \& Loeb,~A. 2011, Phys.\ Rev.\ D, 84, 024032
%
\bibitem[MacFadyen \& Milosavljevi\'c(2008)]{mm08}
MacFadyen,~I.A., \& Milosavljevi\'c, M., 2008, \apj, 672,83
%
\bibitem[Mii \& Totani(2005)]{mt05}
Mii,~H., \& Totani,~T. 2005, \apj, 628, 873
%
\bibitem[Blanchet(2006)]{lb06}
Luc Blanchet. 2006, Living Rev. Relativity, 9, 4
%
\bibitem[Milosavljevi\'c \& Phinney(2005)]{mp05}
Milosavljevi\'c,~M., \& Phinney,~E.S. 2005, \apj, 622, L93
%
\bibitem[Narayan(2000)]{nr00}
Narayan,~R. 2000, \apj, 536, 663
%
\bibitem[Nelemans \& Van den Heuvel(1999)]{nv99}
Nelemans,~G.T.M., \& Van den Heuvel,~E.P.J. 1999, \aap, 352, L87
%
\bibitem[Peters(1964)]{p64}
Peters,~P.C. 1964, Phys.\ Rev., 136, 1224
%
\bibitem[Phinney(2003)]{p03}
Phinney,~E.S., et al. 2003, {\it Big Bang Observer Mission Concept Study} (NASA)
%
\bibitem[Farmer \& Phinney(2003)]{fp03}
Farmer,~A.J., \& Phinney,~E.S. 2003, \mnras, 346, 1197 
%
\bibitem[Sanders et al.(1984)]{sa84}
Sanders,~D.B., Solomon, P.M. \& Scoville,~N.Z. 1984, \apj, 276, 182
%
\bibitem[Seto et al.(2001)]{seto01}
Seto,~N., Kawamura,~S., \& Nakamura,~T. 2001, Phys.\ Rev.\ Lett, 87, 221103
%
\bibitem[Shi and Krolik et al.(2011)]{sklh11}
Shi,~Ji-M., \&  Krolik,~H.J., Lubow,~S.H.,  \& Hawley, F.~J. 2011, arXiv:1110.4866
%
\bibitem[Shakura \& Sunyaev(1973)]{ss73}
Shakura, N.~I., \& Sunyaev, R. A. 1973, \aap, 24, 337
%
\bibitem[Schutz(1986)]{schutz86}
Schutz,~B.F. 1986, \nat, 323, 310
%
\bibitem[Tanaka \& Menou(2010)]{tk10}
Tanaka,~T., \& Menou,~K. 2010, \apj, 714, 404
%
\bibitem[Yagi et al.(2011)]{yagi11}
Yagi,~K., Tanahashi,~N., \& Tanaka,~T. 2011, Phys.\ Rev.\ D, 83, 084036 
%
\bibitem[Yunes et al.(2010)]{yunesG}
Yunes,~N., Pretorius,~F., \& Spergel,~D. 2010, Phys.\ Rev.\  D, 81, 064018
%
\bibitem[Yunes et al.(2011a)]{yunes-perturber}
Yunes,~N., Coleman,~M., \& Thornburg,~J. 2011, Phys.\ Rev.\  D, 83, 044030
%
\bibitem[Yunes et al.(2011b)]{yunes11}
Yunes,~N., Kocsis,~B., Loeb~A., \& Haiman,~Z. 2011, Phys.\ Rev. \ Lett, 107, 171103
%
\end{thebibliography}
\end{document}